\documentclass[twocolumn,showpacs,preprintnumbers,amsmath,amssymb,
superscriptaddress,groupedaddress,pre]{revtex4}

\usepackage{graphicx}
\usepackage{dcolumn}
\usepackage{amsmath}
\usepackage{amssymb}
\usepackage{bm}
\usepackage{epsfig}
\usepackage{color}
\usepackage{multirow}

\begin{document}

\title{Nonlinear waves in disordered chains: probing the limits of
  chaos and spreading}

\author{J.~D.~Bodyfelt} \affiliation{Max Planck Institute for the
  Physics of Complex Systems, N\"othnitzer Str.~38, D-01187 Dresden,
  Germany}

\author{T.~V.~Laptyeva} \affiliation{Max Planck Institute for the
  Physics of Complex Systems, N\"othnitzer Str.~38, D-01187 Dresden,
  Germany}

\author{Ch.~Skokos}\email{hskokos@pks.mpg.de}\thanks{Corresponding
  author.}\affiliation{Max Planck Institute for the Physics of Complex
  Systems, N\"othnitzer Str.~38, D-01187 Dresden, Germany}
\affiliation{Center for Research and Applications of Nonlinear
  Systems, University of Patras, GR-26500, Patras, Greece}

\author{D.~O.~Krimer} \affiliation{Max Planck Institute for the
  Physics of Complex Systems, N\"othnitzer Str.~38, D-01187 Dresden,
  Germany} \affiliation{Theoretische Physik, Universit\"at T\"ubingen,
  D-72076 T\"ubingen, Germany}

\author{S.~Flach} \affiliation{Max Planck Institute for the Physics of
  Complex Systems, N\"othnitzer Str.~38, D-01187 Dresden, Germany}

\date{\today}

\begin{abstract}
  We probe the limits of nonlinear wave spreading in disordered chains
  which are known to localize linear waves. We particularly extend
  recent studies on the regimes of strong and weak chaos during
  subdiffusive spreading of wave packets [EPL {\bf 91}, 30001 (2010)]
  and consider strong disorder, which favors Anderson localization.
  We probe the limit of infinite disorder strength and study
  Fr\"ohlich-Spencer-Wayne models.  We find that the assumption of
  chaotic wave packet dynamics and its impact on spreading is in
  accord with all studied cases. Spreading appears to be asymptotic,
  without any observable slowing down.  We also consider chains with
  spatially inhomogeneous nonlinearity which give further support to
  our findings and conclusions.
\end{abstract}

\maketitle

\section{Introduction}
\label{sec:intro}

Anderson localization \cite{A58} was discovered 50 years ago in
disordered crystals as an accumulation of single particle electronic
wavefunctions and can be interpreted as an interference effect between
multiple scatterings of the electron by random defects of the
potential. As a consequence eigenstates are no longer spatially
extended, but exponentially localized. Anderson localization is a
universal phenomenon of wave physics, unrestricted to quantum
mechanics. Experimental observations were made in noninteracting
Bose-Einstein condensates (BEC) expanding in random optical potentials
\cite{Billy,Roati}, light propagation in spatially random nonlinear
optical media \cite{Exp, Exp2}, and in microwave cavities filled with
randomly distributed scatterers \cite{microwave}. Anderson
localization is a linear wave effect, i.e.~it is well-established for
wave equations which are linear in the wave amplitude.  However, in
many cases one is confronted with a nonlinear response of the
wave-carrying medium; for instance, large light intensities induce a
nonlinear response of the optical medium.  Electron-electron and
electron-phonon interactions also result in substantial deviations
from Anderson localization in solids. In experiments of Bose-Einstein
condensates the interatomic interactions are always present, although
they can be diminished by either decreasing atomic densities or by
exploiting magnetically tunable Feshbach resonances.

From a mathematical perspective, a linear wave equation is integrable,
with each normal mode evolving independently in time.  A localized
wave packet in the presence of Anderson localization will therefore
stay localized as time evolves.  Nonlinearity will usually destroy the
integrability of a system and induce mode-mode interactions. It was
observed numerically that wave packets in such nonlinear disordered
wave equations delocalize in time without respecting Anderson
localization limits \cite{mm98,PS08,FKS09,B11,BCLSB11}.  Thus, there
are several intriguing questions which have attracted much attention
during the last few years: (i) will Anderson localization be destroyed
by arbitrary small strength of nonlinearity or is there a threshold
below which the localization is restored? (ii) will wave packet
spreading, if observed, last forever or will it stop at certain
(though probably very large) time? (iii) is the shape of the initial
wave packet crucial for the details of spreading?  We will mainly
address question (ii) here.

Johansson et al.~\cite{JKA10} conjectured that spreading must
eventually stop and dynamics will become close to regular, assuming
that in these limits the Kolmogorov-Arnold-Moser (KAM) theorem is
applicable, i.e.~that for small wave density regular nonergodic phase
space structures predominate and the dynamics develops along KAM
tori. Other attempts consist in a numerical scaling analysis, in order
to predict and extend results beyond computational ability
\cite{PF10}.  Analytical studies perform perturbation theory to higher
order by treating the strength of nonlinearity as a small parameter
\cite{FKS_Is_09}, conflicting with the explosive growth of secular
terms in higher orders of perturbation theory.  This theory states
that for the disordered discrete nonlinear Schr\"odinger model with
nonlinearity strength exceeding a finite threshold, any initial
localized wave packet cannot fully spread to zero amplitudes at
infinite time. In this case, a part of the excitation is selftrapped
as a result of nonlinearity induced frequency shifts, which tune a
localized excitation out of resonance with its surrounding non-excited
linear modes. However, even in the case of strong nonlinearity,
subdiffusion of the non-selftrapped part is observed
\cite{FKS09}. When strong nonlinearities are avoided numerical studies
showed a rather universal asymptotic subdiffusive spreading of initial
single site excitations \cite{PS08,FKS09,SKKF09}, which is
characterized by a growth of the second moment of the wave packet as
$t^{\alpha}$, with $\alpha \leq 1$ \cite{FKS09,SKKF09}. The
selftrapping theorem \cite{KKFA08} holds irrespective of the strength
of disorder, therefore it is reflecting the properties of a strongly
nonlinear lattice wave equation, rather than peculiarities of waves
propagating in disordered media. Also the selftrapping theorem is
crucially depending on the presence of at least two integrals of
motion, and fails for most nonlinear wave equations with only one
integral of motion.

In \cite{FKS09} the observed wave packet spreading was assumed to be
due to an incoherent excitation of the wave packet exterior, induced
by the chaotic dynamics of the wave packet interior.  The number of
resonant modes in the packet was estimated by considering from
quadruplet and triplet mode-mode interactions \cite{KF10}. A
generalization to higher dimensions $D$ and different nonlinearity
powers were performed. This led to a quantitative prediction for the
subdiffusive wave packet spreading characteristics $\alpha$
\cite{FKS09}.  Its validity was confirmed numerically in
\cite{FKS09,SKKF09,SF10}.  Recently, it has been predicted
theoretically \cite{F10,LBKSF10} and verified numerically
\cite{LBKSF10} that a potentially long-lasting strong chaos regime
induces faster (though still subdiffusive) spreading, which is
followed by the asymptotic and slower weak chaos subdiffusive
spreading.  Notably, published numerical data did not report on a
further slowing down of spreading, when starting from the weak chaos
regime.

In this paper we present results of extensive numerical studies of
wave packet spreading in various models of disordered nonlinear
one-dimensional lattices. In particular, we consider different initial
excitations and scan the parameter space of disorder strength and
nonlinearity over a wide region. The main aim is to test the
applicability of previously derived spreading laws, and to search for
indications of a continuation of the weak chaos spreading, or for
indications of a slowing down, as conjectured by others.

\section{Models}
\label{sec:all_models}

\subsection{Discrete Nonlinear Schr\"odinger and Klein-Gordon chains}
\label{sec:models}

In our study we consider various one-dimensional lattice models. The
first one is the disordered discrete nonlinear Schr\"odinger equation
(DNLS) described by the Hamiltonian function
\begin{equation}
\mathcal{H}_{D}= \sum_{l} \epsilon_{l} |\psi_{l}|^2+\frac{\beta}{2}
|\psi_{l}|^{4} - (\psi_{l+1}\psi_l^* +\psi_{l+1}^* \psi_l),
\label{RDNLS}
\end{equation}
in which $\psi_{l}$ are complex variables, $l$ are the lattice site
indices and $\beta \geq 0$ is the nonlinearity strength.  The random
on-site energies $\epsilon_{l}$ are chosen uniformly from the interval
$\left[-\frac{W}{2},\frac{W}{2}\right]$, with $W$ denoting the
disorder strength.  The equations of motion are generated by
$\dot{\psi}_{l} = \partial \mathcal{H}_{D}/ \partial (i
\psi^{\star}_{l})$:
\begin{equation}
i\dot{\psi_{l}}= \epsilon_{l} \psi_{l} +\beta |\psi_{l}|^{2}\psi_{l}
-\psi_{l+1} - \psi_{l-1}\;.
\label{RDNLS-EOM}
\end{equation}
This above set of equations conserves both the energy of
Eq.~(\ref{RDNLS}), and the norm $S = \sum_{l}|\psi_l|^2$.

The second model we consider is the quartic Klein-Gordon (KG) lattice,
given as
\begin{equation}
\mathcal{H}_{K}= \sum_{l} \frac{p_{l}^2}{2} + \frac{\tilde{\epsilon}_{l}}{2}
u_{l}^2 + \frac{1}{4} u_{l}^{4}+\frac{1}{2W}(u_{l+1}-u_l)^2,
\label{RQKG}
\end{equation}
where $u_l$ and $p_l$ respectively are the generalized coordinates and
momenta on site $l$, and $\tilde{\epsilon}_{l}$ are chosen uniformly
from the interval $\left[\frac{1}{2},\frac{3}{2}\right]$.  The
equations of motion are $\ddot{u}_{l} = - \partial \mathcal{H}_{K}
/\partial u_{l}$ and yield
\begin{equation}
\ddot{u}_{l} = - \tilde{\epsilon}_{l}u_{l} -u_{l}^{3} + \frac{1}{W}
(u_{l+1}+u_{l-1}-2u_l)\;.
\label{KG-EOM}
\end{equation}
This set of equations only conserves the energy of
Eq.~(\ref{RQKG}). The scalar measure of energy resulting from
Eq.~(\ref{RQKG}) we shall henceforth label as $H$. This scalar value
$H \geq 0$ serves as a control parameter of nonlinearity, similar to
$\beta$ for the DNLS case.

For $\beta=0$ and $\psi_{l} = A_{l} \exp(-i\lambda t)$,
Eq.~(\ref{RDNLS-EOM}) reduces to the linear eigenvalue problem
\begin{equation}
\lambda A_{l} = \epsilon_{l} A_{l} - A_{l-1}-A_{l+1}\;.
\label{EVequation}
\end{equation}
The normalized eigenvectors $A_{\nu,l}$ ($\sum_l A_{\nu,l}^2=1)$ are
the corresponding normal modes (NMs), and the eigenvalues
$\lambda_{\nu}$ are the frequencies of these NMs.  The width of the
eigenfrequency spectrum $\lambda_{\nu}$ in Eq.~(\ref{EVequation}) is
$\Delta_D=W+4$ with $\lambda_{\nu} \in \left[ -2 -\frac{W}{2}, 2 +
  \frac{W}{2} \right] $. The coefficient $1/ (2W)$ in Eq.~(\ref{RQKG})
was chosen so that the linear parts of the Hamiltonians,
Eqs.~(\ref{RDNLS},\ref{RQKG}) would correspond to the same eigenvalue
problem. In the limit $H \rightarrow 0$ (in practice by neglecting the
nonlinear term $u_l^4/4$) the KG model of Eq.~(\ref{RQKG}) - with
$u_{l} = A_{l} \exp(i\omega t)$ - is reduced to the same linear
eigenvalue problem of Eq.~(\ref{EVequation}), under the substitutions
$\lambda = W\omega^2-W -2$ and $\epsilon_l=W(
\tilde{\epsilon}_{l}-1)$.  The width of the squared frequency
$\omega_{\nu}^2$ spectrum is $\Delta_{K}= 1+ \frac{4}{W}$ with
$\omega_{\nu}^2 \in \left[ \frac{1}{2},\frac{3}{2} +
  \frac{4}{W}\right] $.  Note that $\Delta_D = W \Delta_{K}$. As in
the case of DNLS, $W$ determines the disorder strength.

The asymptotic spatial decay of an eigenvector is given by $A_{\nu,l}
\sim {\rm e}^{-l/\xi(\lambda_{\nu})}$ where $\xi(\lambda_{\nu})$ is
the localization length. In the case of weak disorder, $W \rightarrow
0$, the localization length is approximated \cite{KK93,KF10} as
$\xi(\lambda_{\nu}) \leq \xi(0) \approx 100/W^2$. The NM participation
number $p_{\nu} = 1/\sum_l A_{\nu,l}^4$ characterizes the spatial
extend of the NM. An average measure of this extent is the
localization volume $V$, which is of the order of $3.3 \xi(0)\approx
330/W^2$ for weak disorder and unity in the limit of strong disorder,
$W \rightarrow \infty$ \cite{KF10}. The average spacing of eigenvalues
of NMs within the range of a localization volume is then $d\approx
\Delta/ V$, with $\Delta$ being the spectrum width.  The two frequency
scales $d \leq \Delta$ determine the packet evolution details in the
presence of nonlinearity.

In order to write the equations of motion of Hamiltonian (\ref{RDNLS})
in the normal mode space of the system we insert $\psi_l=\sum_{\nu}
A_{\nu,l} \phi_{\nu}$ in (\ref{RDNLS-EOM}), with $|\phi_{\nu}|^2$
denoting the time-dependent amplitude of the $\nu$th NM. Then, using
Eq.~(\ref{EVequation}) and the orthogonality of NMs the equations of
motion (\ref{RDNLS-EOM}) read
\begin{equation}
  i \dot{\phi}_{\nu} = \lambda_{\nu} \phi_{\nu} + \beta
  \sum_{\nu_1,\nu_2,\nu_3} I_{\nu,\nu_1,\nu_2,\nu_3} \phi^*_{\nu_1}
  \phi_{\nu_2} \phi_{\nu_3}\;
\label{NMeq}
\end{equation}
with the overlap integral 
\begin{equation}
I_{\nu,\nu_1,\nu_2,\nu_3} = \sum_{l} A_{\nu,l} A_{\nu_1,l} A_{\nu_2,l}
A_{\nu_3,l}\;.
\label{OVERLAP}
\end{equation}
The frequency shift of a single site oscillator induced by the
nonlinearity is $\delta_{l} = \beta |\psi_l|^{2}$ for the DNLS
model. The squared frequency shift of a single site oscillator induced
by the nonlinearity for the KG system is $\delta_{l} = (3 E_l)/(2
\tilde{\epsilon}_l)$, with $E_l$ being the energy of the
oscillator. Since all NMs are exponentially localized in space, each
effectively couples to a finite number of neighbor modes.  The
nonlinear interactions are thus of finite range; however, the strength
of this coupling is proportional to the norm (energy) density for the
DNLS (KG) model. If the packet spreads far enough, we can generally
define two norm (energy) densities: one in real space, $n_l
=|\psi_l|^2$ ($E_l$) and the other in NM space, $n_{\nu}
=|\phi_{\nu}|^2$ ($E_{\nu}$). Averaging over realizations, no strong
difference is seen between the two, and therefore, we treat them
generally as some characteristic norm ($n$) or energy ($E$)
density. The frequency shift due to nonlinearity is then $\delta_D
\sim \beta n$ for the DNLS model, while the squared frequency shift is
$\delta_K \sim 3 E/2$ for the KG lattice. The basic characteristics of
both models are summarized in Table \ref{tab:char}.

\begin{table*} 
\centering
{
\renewcommand{\arraystretch}{2}
\begin{tabular}{l||c|c|}
  \cline{2-3} & DNLS & KG \\ \hline \hline \multicolumn{1}{|l||}{On-site
    energies} & $\displaystyle \epsilon_l \in
  \left[-\frac{W}{2},\frac{W}{2}\right]$ & $ \displaystyle \tilde{\epsilon}_{l}
  \in \left[\frac{1}{2},\frac{3}{2}\right]$ \\ \hline
  \multicolumn{1}{|l||}{Spectrum} & $ \displaystyle \lambda_{\nu} \in \left[ -2
    -\frac{W}{2}, 2 + \frac{W}{2} \right] $ & $ \displaystyle \omega_{\nu}^2 \in
  \left[ \frac{1}{2},\frac{3}{2} + \frac{4}{W}\right] $ \\ \hline
  \multicolumn{1}{|l||}{Spectrum width $\Delta$} & $\Delta_D=W+4$ &
  $\displaystyle \Delta_K=\frac{W+4}{W}$ \\ \hline
  \multicolumn{1}{|l||}{\multirow{2}{*}{Localization volume $V$ $\left\lbrace
\begin{array}{c}
W \rightarrow 0 \,\,\, \mbox{} \\
W \rightarrow \infty \,\,\, \mbox{} 
\end{array} \right.$ }} & \multicolumn{2}{c|}{$\displaystyle V=\frac{330}{W^2}$} \\ \cline{2-3} 
\multicolumn{1}{|l||}{}& \multicolumn{2}{c|}{$\displaystyle V\sim 1$} \\
\hline \multicolumn{1}{|l||}{\multirow{2}{*}{Average spacing $d$
    $\left\lbrace
\begin{array}{c}
W \rightarrow 0 \,\,\, \mbox{} \\
W \rightarrow \infty \,\,\, \mbox{} 
\end{array} \right.$ }} & $\displaystyle d_D \sim W^2$ & $\displaystyle d_K \sim W$ \\ \cline{2-3}
\multicolumn{1}{|l||}{} & $\displaystyle d_D \sim W$ & $\displaystyle d_K \sim
\mbox{const.}$ \\ \hline \multicolumn{1}{|l||}{Nonlinear energy shift
$\delta$} & $\displaystyle \delta_D \sim \beta n$ & $\displaystyle \delta_K
\sim \frac{3}{2} E$ \\ \hline
\end{tabular}
}
\caption{Characteristic quantities of the DNLS (\ref{RDNLS}) and the KG
  (\ref{RQKG}) models. The dependence on the strength of disorder $W$ of both
  the localization volume $V$ and of the average spacing $d$ of NMs' eigenvalues
  within the range of $V$, is given for the limiting cases of weak $W
  \rightarrow 0$ and strong disorder $W \rightarrow \infty$. Note that $n$ ($E$)
  represents a general characteristic norm (energy) of wave packets of the DNLS
  (KG) model. }
\label{tab:char}
\end{table*}

We order the NMs in space by increasing value of the center-of-norm
coordinate $X_{\nu}=\sum_l l A_{\nu,l}^2$
\cite{FKS09,SKKF09,SF10,LBKSF10}.  For DNLS we follow normalized norm
density distributions $z_{\nu}\equiv |\phi_{\nu}|^2/\sum_{\mu}
|\phi_{\mu}|^2$, while for KG we follow normalized energy density
distributions $z_{\nu}\equiv E_{\nu}/\sum_{\mu} E_{\mu}$ with $E_{\nu}
= \dot{A}^2_{\nu}/2+\omega^2_{\nu}A_{\nu}^2/2$, where $A_{\nu}$ is the
amplitude of the $\nu$th NM and $\omega^2_\nu$ its squared
frequency. We measure the second moment $m_2= \sum_{\nu}
(\nu-\bar{\nu})^2 z_{\nu}$ (with $\bar{{\nu}} = \sum_{\nu} \nu
z_{\nu}$), which quantifies the wave packet's spreading width; the
participation number $P=1 / \sum_{\nu} z_{\nu}^2$, i.e. the number of
the strongest excited modes in $z_{\nu}$; and the compactness index
$\zeta=P^2/m_2$, which quantifies the inhomogeneity of a wave
packet. Thermalized distributions have $\zeta \approx 3$, while $\zeta
\ll 3$ indicates very inhomogeneous packets, e.~g.~sparse (with many
holes) or partially selftrapped ones (see \cite{SKKF09} for more
details).  In addition, following Anderson's definition of
localization \cite{A58}, we measure the fraction $S_V$ ($H_V$) of the
wave packet norm (energy) in a localization volume $V$ around the
initially excited state in real space.  For a localized state this
fraction asymptotically tends to a constant nonzero value, while it
goes to zero in the case of delocalization.

\subsection{Fr\"ohlich-Spencer-Wayne chain}
\label{sec:FSW_model}

In the limit of strong disorder ($W\rightarrow \infty$) the DNLS and
KG models suffer from increasing computational times needed to observe
any nontrivial dynamics. This is because the eigenvectors tend to
single site profiles, i.e.~the overlap integrals become very small.
Fr\"ohlich, Spencer and Wayne (FSW) suggested considering a modified
Hamiltonian, which operates directly in normal mode space for the
strong disorder limit, but considers artificial rescaled anharmonic
interactions between neighboring NMs in order to rescale time
\cite{FSW86}:
\begin{equation}
  \mathcal{H}= \sum_{\nu} \frac{p_{\nu}^2}{2} + \frac{\epsilon_{\nu}}{2} u_{\nu}^2
  +\frac{1}{4}(u_{\nu+1}-u_\nu)^4,
\label{FSW}
\end{equation}
where the NMs are equivalent to the single site oscillators. The NM
eigenvalues $\epsilon_{\nu}$ are considered to be uncorrelated, also
for nearest neighbors. This is different from the DNLS and KG
models. Also the FSW chain has only pair interactions between NMs
(sites).  Note also that the nonlinear part of the FSW Hamiltonian is
invariant under any shift $u_{\nu} \rightarrow u_{\nu} + a$, as
opposed to the KG model.

\subsection{Models with spatially inhomogeneous nonlinearity}
\label{sec:hybrid_model}

We also consider two variants of DNLS and KG models with spatially
inhomogeneous nonlinearity. The first type of lattices is composed of
linear coupled oscillators except for a central region of length $L$
where nonlinearities are present. We refer to this type as the LNL
(Linear-Nonlinear-Linear) model. The second type is called NLN
(Nonlinear-Linear-Nonlinear) and is the exact counterpart of the
previous one, since the linear part of the lattice is located at the
central $L$ sites.

\section{Wave packet evolution}
\label{sec:evol}

\subsection{Theoretical predictions}
\label{sec:theory}

\subsubsection{DNLS and KG}

The evolution of wave packets in nonlinear disordered chains can be
expected to be selftrapped for strong nonlinearities, or show no
selftrapping for weaker nonlinearities. The existence of the
selftrapping regime was theoretically predicted for the DNLS model in
\cite{KKFA08} (see also \cite{SKKF09} for more details). According to
the theorem stated in \cite{KKFA08}, for large enough nonlinearities
($\delta_D>\Delta_D$) single site excitations cannot uniformly spread
over the entire lattice. Consequently, a part of the wave packet will
remain localized, although the theorem does not prove that the
location of this inhomogeneity is constant in time.

If the nonlinear shift $\delta$ moves the frequencies of some of the
initially excited oscillators out of the linear spectrum, it tunes
them out of resonance and part of the wave packet will be
selftrapped. In our study we consider initial ``block`` wave packets,
where $L$ central oscillators of the lattice are excited having the
same norm (energy). Since we consider many random disorder
realizations (of the order of few hundreds) we expect that, on
average, the linear frequencies of the initially excited lattice sites
$\epsilon_l$ ($\tilde{\epsilon}_{l}$) cover the whole range of
permitted values $\left[-\frac{W}{2},\frac{W}{2}\right]$ ($
\left[\frac{1}{2},\frac{3}{2}\right]$). Thus, some of these
frequencies are tuned out of resonance if $\delta_D \geq 2$ ($\delta_D
\gtrsim 1/W$). These conditions for the possible appearance of
selftrapping are less strict than the theoretically defined ones
\cite{KKFA08,SKKF09} and are, in general, in good agreement with
numerical simulations. In particular, the selftrapping regime was
numerically observed for single-site excitations
\cite{FKS09,SKKF09,SF10} and for extended excitations \cite{LBKSF10},
both for the DNLS and the KG models, despite the fact that the KG
system conserves only the total energy $E$, and the selftrapping
theorem can not be applied there.

When selftrapping is avoided for $\delta_D <2$ ($\delta_K \lesssim
1/W$), two different spreading regimes were predicted, having
different dynamical characteristics: an asymptotic weak chaos regime,
and a potential intermediate strong chaos one \cite{F10}. Numerical
verifications of the existence of these two regimes were presented in
\cite{SF10,LBKSF10}. In the weak chaos regime, for $L\geq V$ and
$\delta < d$ most of the NMs are weakly interacting with each
other. Then the subdiffusive spreading of the wave packet is
characterized by $m_2 \sim t^{1/3}$.  If the nonlinearity is weak
enough to avoid selftrapping, yet strong enough to ensure $\delta >
d$, the strong chaos regime is realized.  Wave packets in this regime
initially spread faster than in the case of weak chaos, with $m_2 \sim
t^{1/2}$. Since the norm density drops with further spreading,
$\delta$ is dropping in time as well, and eventually the wave packet
enters the weak chaos regime, where its evolution is characterized by
slower spreading with $m_2 \sim t^{1/3}$. The wave packet evolution in
both the weak and the strong chaos spreading regimes is also expected
to be characterized by an increase of the participation number as
$P\sim t^{\alpha/2}$ when $m_2 \sim t^{\alpha}$.

Let as now discuss the spreading of wave packets when $L<V$.  The
packet will initially spread over the localization volume $V$ during a
time interval $\tau_{in}\sim2 \pi/d$, even in the absence of
nonlinearities \cite{F10,LBKSF10}.  The initial average norm (energy)
density $n_{in}$ ($E_{in}$) of the wave packet is then lowered to
$n(\tau_{in})\approx n_{in} L/V$ ($E(\tau_{in})\approx E_{in}
L/V$). The further spreading of the wave packet in the presence of
nonlinearities is then determined by these reduced densities. Note
that for single-site excitations ($L=1$), the strong chaos regime
therefore completely disappears and the wave packet evolves either in
the weak chaos or in the selftrapping regimes
\cite{PS08,FKS09,SKKF09,VKF09}.

\subsubsection{FSW chain}
\label{sec:FSW_C}

There are no existing theoretical predictions for wave packet dynamics
in the FSW chain.  The FSW case can be considered as a strong disorder
limit of the KG model, emulating the dynamics of the latter in NM
space. However, in the KG and DNLS case, triplet interactions between
NMs are present and necessary in order to allow for finite (though
small) resonance probabilities for small (but finite) energy and norm
densities \cite{F10,KF10}. Pair interactions will cease to produce NM
resonances for sufficiently small densities due to level repulsion
within one localization volume \cite{SKKF09,F10,KF10}.  The FSW chain
keeps only pair interactions. At the same time, level repulsion
between neighboring (interacting) NMs are absent in the FSW
case. Therefore, a theoretical analysis similar to the DNLS and KG
case \cite{F10} appears to be possible.  Its details will be
considered in a future publication.  The width of the linear spectrum
$\Delta_{FSW}=1$.  The average spacing of nearest neighbor eigenvalues
$d_{FSW} \sim \Delta_{FSW}$.  Therefore, we expect that only the
asymptotic regime of weak chaos, and the regime of selftrapping can be
expected.

\subsubsection{LNL and NLN chains}
\label{sec:LNL_NLN}

The LNL chain can be expected to start in a chaotic wave packet
spreading regime as long as the wave packet is confined mainly to the
finite size N (nonlinear) part.  However, the more the wave packet
spreads, the more it extends into the infinitely extended L (linear)
parts. Resonances and chaos are therefore confined to the finite N
part.  Since distant NMs in the L part are exponentially weakly
interacting with the chaotic NMs in the N part, their excitation - if
at all - will take times which increase exponentially with growing
distance. Therefore, the wave packet will spread (if at all) slower
than any power law.  Thus, the LNL model is the only model we consider
here, where almost trivial slowing down of spreading is expected.

Recently, the dynamics of a similar system was theoretically
investigated in \cite{AS09}, where a disordered subsystem of coupled
anharmonic oscillators, linearly coupled to an infinite harmonic
system, was considered. In this work the conditions which permit the
persistence of the discrete breather of the isolated anharmonic system
for small but nonvanishing couplings to the harmonic lattice were
derived, and cases characterized by energy transfer to the harmonic
system were also discussed.

The NLN chain is expected to behave differently. As long as the wave
packet is confined mainly to the L region, the dynamics is regular,
and no spreading should occur.  For large enough time, some part of
the packet will leak out into the N regions.  Therefore, finally
spreading of the wave packet should occur.

\subsection{Numerical results}
\label{sec:num}

\subsubsection{Methods}
\label{sec:methods}

We consider compact DNLS wave packets at $t =0$ spanning a width $L$
centered in the lattice, such that within $L$ there is a constant
initial norm of $n_{in}=1$ and a random phase at each site , while
outside the width $L$ the norm density is zero. In the KG case, we
excite each site in the width $L$ with the same energy, $E=H/L$,
i.e.~initial momenta of $p_l =\pm \sqrt{2E}$ with randomly assigned
signs.

We use symplectic integration schemes of the SABA family of
integrators \cite{SKKF09,LR01,SG10} for the integration of equations
(\ref{RDNLS-EOM}) and (\ref{KG-EOM}). The particular symplectic scheme
used for the DNLS model is described in the Appendix. The number of
lattice sites $N$ and the integration time step $\tau$ varied between
$N=1000$ to $N=2000$ and $\tau=0.01$ to $\tau=0.1$, in order to
exclude finite size effects in the wave packet evolution, and in order
to reach long integration times up to $10^{7} -10^{9}$ time units with
feasible CPU times. In all our simulations, the relative energy and
norm errors are kept smaller than $10^{-3}$.  For each parameter set
we averaged our data over $1000$ different disorder realizations,
unless otherwise stated, and denote this by $ \langle \ldots \rangle$.
In particular, we compute $m_2$ and $P$, and we smooth $\langle
\log_{10} m_2 \rangle$ and $\langle \log_{10} P \rangle$ with a
locally weighted regression algorithm \cite{CD88}, and then apply a
central finite difference to calculate the local derivatives
\begin{equation}
  \alpha_m=\frac{d \langle \log_{10} m_2 \rangle}{d \log_{10} t}, \,\,\,
  \alpha_P=\frac{d \langle \log_{10} P \rangle}{d \log_{10} t}.
\label{eq:deriv}
\end{equation}

\subsubsection{Weak disorder}
\label{sec:weak}

\begin{figure*}
\includegraphics[scale=0.35,angle=0]{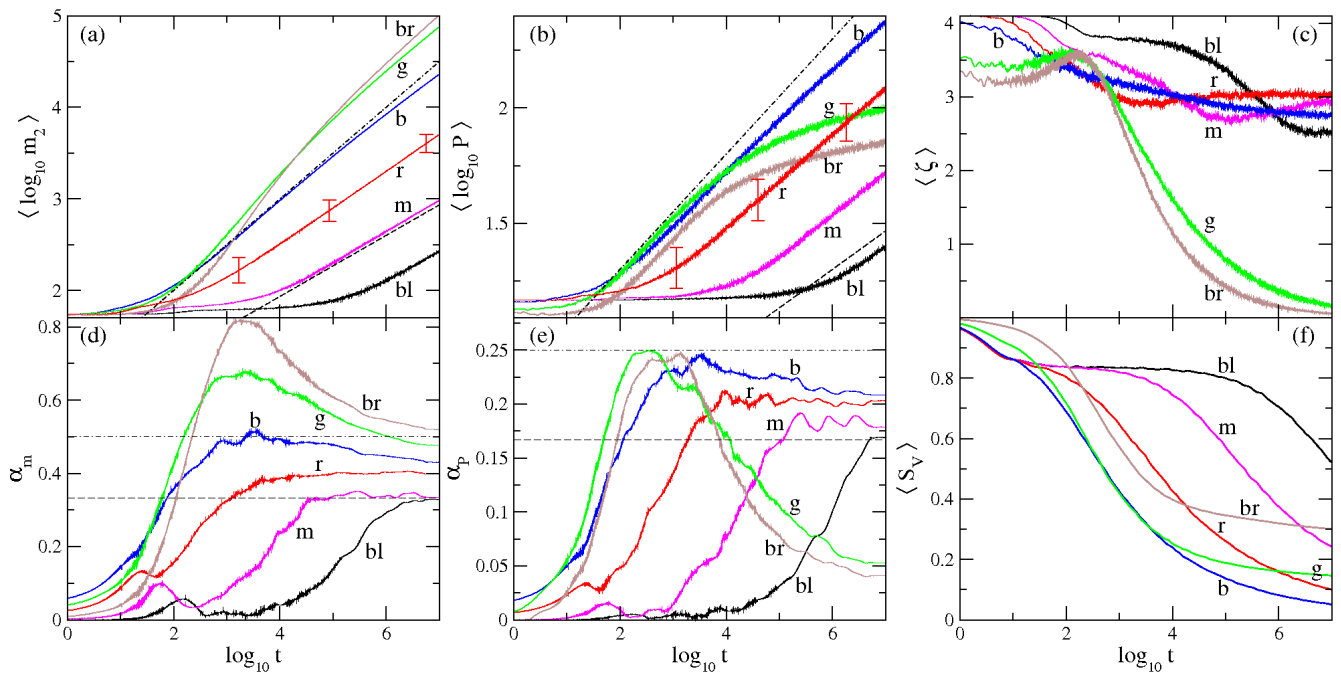} 
\caption{(Color online) DNLS, $W=4$: Evolution of (a) $\langle
  \log_{10}m_2(t) \rangle$, (b) $\langle \log_{10}P(t) \rangle$, (c)
  $\langle \zeta(t) \rangle$, (d) finite-difference derivative
  $\alpha_m (t)$ for the smoothed $m_2$ data of panel (a), (e)
  finite-difference derivative $\alpha_P (t)$ for the smoothed $P$
  data of panel (b), and (f) $\langle S_V(t) \rangle$ for the
  spreading of wave packets with initial width $L=21$ and
  $\beta=0.012, \, 0.04 , \, 0.18, \, 0.72, \, 3.6, \, 8.4$ [(bl)
  black; (m) magenta; (r) red; (b) blue; (g) green; (br) brown].  In
  panels (a), (b), (d) and (e) straight lines correspond to
  theoretically predicted power laws $m_2 \sim t^{\alpha}$, $P\sim
  t^{\alpha/2}$ with $\alpha=1/3$ (dashed lines) and $\alpha=1/2$
  (dotted lines). Error bars in panels (a) and (b) denote
  representative standard deviation errors.}
\label{fig_DNLS_4}
\end{figure*}

In this subsection we considerably extend the reports on the
observation of weak chaos, strong chaos, the crossover between both,
and the selftrapping regime in Ref. \cite{LBKSF10}.  In
Fig.~\ref{fig_DNLS_4} we show results for the DNLS model with $W=4$
and $L=V=21$, for six different values of the nonlinearity strength
$\beta$.  The time evolution of $\langle \log_{10}m_2(t) \rangle$ and
$\langle \log_{10} P(t) \rangle$ is plotted in
Figs.~\ref{fig_DNLS_4}(a) and \ref{fig_DNLS_4}(b), respectively. The
evolution of the compactness index $\langle \zeta(t) \rangle$ is shown
in Fig.~\ref{fig_DNLS_4}(c). In Figs.~\ref{fig_DNLS_4}(d) and
\ref{fig_DNLS_4}(e) we plot the time dependence of the numerically
computed derivatives $\alpha_m (t)$ and $\alpha_P (t)$
(\ref{eq:deriv}) obtained from the smoothed curves of
Figs.~\ref{fig_DNLS_4}(a) and \ref{fig_DNLS_4}(b)
respectively. Finally, in Fig.~\ref{fig_DNLS_4}(f) the values of
$\langle S_V (t) \rangle$ are plotted.

The weak chaos dynamics is observed in Fig.~\ref{fig_DNLS_4} for
$\beta=0.012$ (black curves) and $\beta=0.04$ (magenta curves).
Initially, the wave packets remain localized and all quantities of
Fig.~\ref{fig_DNLS_4} are constant with $\alpha_m$, $\alpha_P$ being
practically zero. After some detrapping time $t_d$ the wave packets
start to subdiffusively spread with $m_2 \sim t^{1/3}$ and $P \sim
t^{1/6}$ (Figs.~\ref{fig_DNLS_4}(a), (b), (d) and (e)). In addition,
the compactness index $\langle \zeta \rangle \approx 3$
(Fig.~\ref{fig_DNLS_4}(c)), indicating that wave packets are well
thermalized inside. The tendency towards complete delocalization of
wave packets is clearly depicted in the evolution of the averaged norm
fraction $\langle S_V \rangle$ which remains at the $L=21$ initially
excited sites (Fig.~\ref{fig_DNLS_4}(f)). After the detrapping time
$t_d$ $\langle S_V \rangle$ decreases continuously up to the final
integration time $t=10^7$.

Increasing the value to $\beta=0.18$, the initial spreading dynamics
enters the crossover between weak and strong chaos, and a faster
spreading is observed.  Spreading sets in earlier, the compactness
index again indicates thermalized wave packets, and the local
derivatives $\alpha_m$ and $\alpha_P$ increase up to 0.4 and 0.2
respectively, with a possibly very slow decreasing at even larger
times.  $\langle S_V \rangle$ again continuously decreases to zero
indicating complete delocalization.

For $\beta=0.72$, we fully enter the strong chaos regime.  Most
importantly we observe a saturation of the local exponent $\alpha_m$
around the theoretical value 1/2, with a subsequent decay, again as
predicted by theory \cite{F10}, and first observed in \cite{LBKSF10}.

Finally, for $\beta=3.6$ and $\beta=8.4$ (green and brown curves
respectively in Fig.~\ref{fig_DNLS_4}) the dynamics enters the
selftrapping regime.  We observe that a part of a wave packet remains
localized, while the remainder spreads.  The spreading portion results
in a continuous increase of $m_2$ (Fig.~\ref{fig_DNLS_4}(a)) which
initially is characterized by large values of $\alpha_m > 1/2$
(Fig.~\ref{fig_DNLS_4}(d)). For larger time $\alpha_m$ decreases below
1/2.  The evolution appears to be rather complex, and is not captured
by the theoretical considerations in \cite{F10}. The large values of
$\alpha_m$ may be due to temporal trapping and detrapping processes in
this strongly nonequilibrium dynamics of the wave packet, with a final
trapped packet fraction remaining. Therefore, we observe that $\langle
\log_{10} P \rangle$ starts to level off (Fig.~\ref{fig_DNLS_4}(b)),
$\alpha_P$ tends to very small values (Fig.~\ref{fig_DNLS_4}(e)) and
$\langle \zeta \rangle$ tends to zero (Fig.~\ref{fig_DNLS_4}(c)). In
Fig.~\ref{fig_DNLS_4}(f), we observe that the values of $\langle S_V
\rangle$ saturate to higher values for $\beta=8.4$.

\begin{figure*}
\includegraphics[scale=0.35,angle=0]{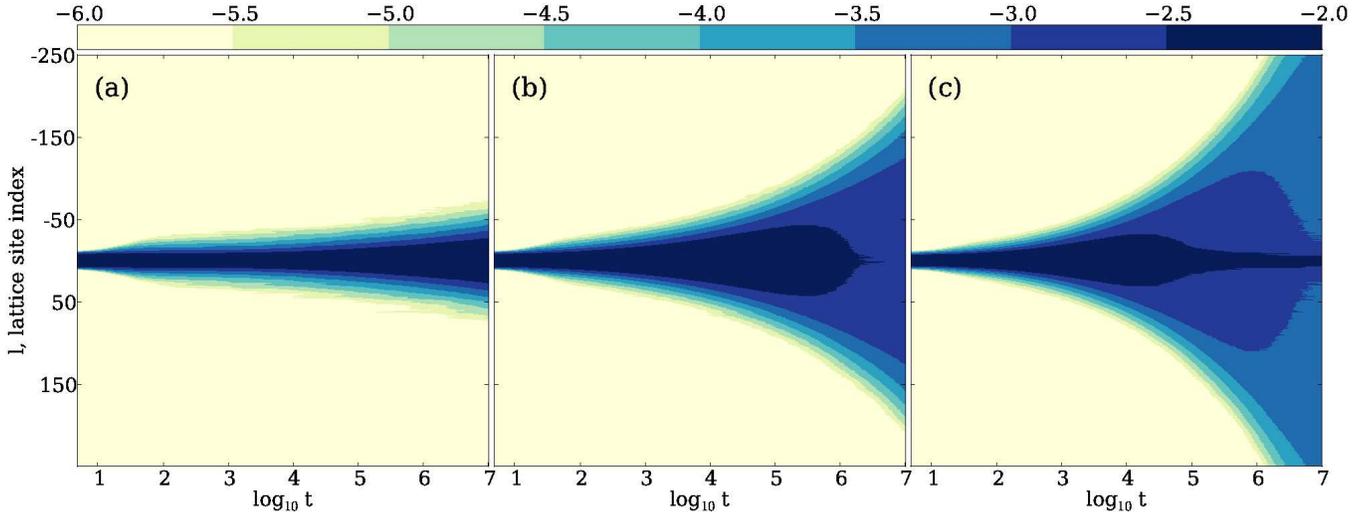} 
\caption{(Color online) DNLS, $W=4$: Time evolution of average norm
  density distributions $\langle z_l \rangle $ in real space for (a)
  $\beta=0.04$, (b) $\beta=0.72$ and (c) $\beta=3.6$. The color scales
  shown on top of panels (a)-(c) are used for coloring each lattice
  site according to its $\log_{10} \langle z_l \rangle$ value.}
\label{fig_DNLS_KG_W4_profiles}
\end{figure*}

For these typical cases of weak chaos ($\beta=0.04$), strong chaos
($\beta=0.72$) and selftrapping ($\beta=3.6$) we present in
Figs.~\ref{fig_DNLS_KG_W4_profiles}(a)-(c) the time evolution of the
averaged norm density distributions $\langle z_l \rangle $ in real
space.  All simulations presented in
Fig.~\ref{fig_DNLS_KG_W4_profiles} started from the same initial
profile with size $L=V$, therefore the width of the localization
volume set by the linear case is the width of the distributions at the
shortest times in the plots.  In the weak chaos regime
(Fig.~\ref{fig_DNLS_KG_W4_profiles}(a)), the wave packets remain close
to their initial configuration for some times, as the high $\langle
z_l \rangle $ values in the region of the initial excitation indicate,
followed by delocalization at larger times. Thus, at $t=10^7$ the
averaged wave packet has spread to about $300$ sites with $\langle z_l
\rangle > 10^{-5}$.  Therefore, the wave packet spreads continuously
over distances which are an order of magnitude larger than the limits
set by the linear theory and destroying Anderson localization.  In the
case of strong chaos (Fig.~\ref{fig_DNLS_KG_W4_profiles}(b)),
spreading is even faster, leading to more extended profiles at
$t=10^7$: about $700$ sites with $\langle z_l \rangle > 10^{-5}$.  In
the selftrapping regime (Fig.~\ref{fig_DNLS_KG_W4_profiles}(c)), the
spreading part of the wave packet covers $1500$ sites, with another
clearly visible part staying selftrapped at the initial excitation
region.  The curved fronts in the density plots in
Fig.~\ref{fig_DNLS_KG_W4_profiles} follow from the theoretical
prediction $m_2 \sim t^{\alpha}$ with which leads to a packet width
$\mathcal{N}\sim\sqrt{m_2}$ and $\mathcal{N} \sim {\rm e}^{(\alpha
  \log_{10} t) /2}$.

\begin{figure}
\includegraphics[width=\columnwidth,angle=0]{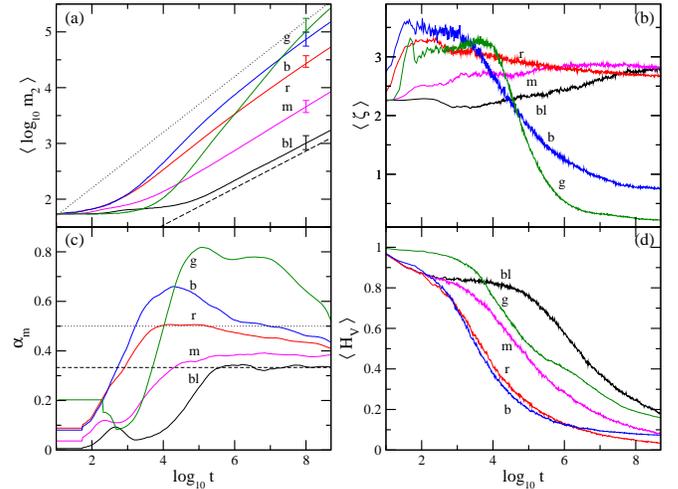} 
\caption{ (Color online) KG, $W=4$: Evolution of (a) $\langle
  \log_{10}m_2(t) \rangle$, (b) $\langle \zeta(t) \rangle$, (c)
  $\alpha_m (t)$, and (d) $\langle H_V(t) \rangle$ versus $\log_{10}t$
  for the spreading of initially compact wave packets of width $L=21$
  with $E=0.01, \, 0.04 , \, 0.2, \, 0.75, \, 3$ [(bl) black; (m)
  magenta; (r) red; (b) blue; (g) green]. In panels (a), and (c)
  straight lines correspond to the theoretically predicted power laws
  $m_2 \sim t^{\alpha}$ with $\alpha=1/3$ (dashed lines) and
  $\alpha=1/2$ (dotted lines). Error bars in panel (a) denote
  representative standard deviation errors.}
\label{fig_KG_4}
\end{figure}

For the KG model (\ref{RQKG}) with $W=4$ we present in
Fig.~\ref{fig_KG_4} similar results to the ones for the DNLS
model. For small values of the initial energy density $E=0.01$, the
characteristics of the weak chaos regime are observed: $m_2 \sim
t^{1/3}$ (black curve in Fig.~\ref{fig_KG_4}(a)) after a detrapping
time $t_d\approx 10^5$, wave packets remain compact as they spread
since $\langle \zeta \rangle \approx 3$ (Fig.~\ref{fig_KG_4}(b)), and
the fraction $\langle H_V \rangle$ of the energy of the initially
$L=21$ excited sites decreases (Fig.~\ref{fig_KG_4}(d)).  For $E=0.04$
we enter the crossover region between the weak and the strong chaos
regimes, with characteristics similar to the DNLS case.  For $E=0.2$
(red curves in Fig.~\ref{fig_KG_4}) we observe the typical behavior of
the strong chaos scenario: spreading is characterized by a saturated
$\alpha_m \approx 1/2$ (Fig.~\ref{fig_KG_4}(c)) for about two decades
($\log_{10} t \approx 3.5- 5.5$), followed by a crossover to the weak
chaos dynamics with $\alpha_m$ decreasing. Getting closer to the
selftrapping regime for $E=0.75$ (blue curves in Fig.~\ref{fig_KG_4}),
or being deep inside it for $E=3$ (green curves in
Fig.~\ref{fig_KG_4}) the characteristics of the selftrapping behavior
appear, since $\langle \zeta \rangle $ decreases
(Fig.~\ref{fig_KG_4}(b)), and $\langle H_V \rangle $ tends to
stabilize to non-zero small values (Fig.~\ref{fig_KG_4}(d)). Similar
to the $\beta=3.6$ and $\beta=8.4$ cases of the DNLS model, the
evolution of $m_2$ (Fig.~\ref{fig_KG_4}(a)) shows an initial phase of
fast growth with $\alpha_m > 1/2$ (Fig.~\ref{fig_KG_4}(c)) followed by
a lowering in the values of $\alpha_m$.

Our numerical results are in accord with the predictions of weak and
strong chaos regimes, as well as of the crossover from strong to weak
chaos.  Selftrapping is observed as well, with less understood
strongly nonequilibrium dynamics of the trapped and spreading packet
parts.  We vehemently stress that in all our simulations we never
observed any evidence of a wave packet transition from the weak chaos
regime, characterized by $\alpha_m=1/3$, to a subsequent slowing down
of spreading, which would lead to $\alpha_m < 1/3$.

\subsubsection{Strong disorder}
\label{sec:strong}

\begin{figure}
\includegraphics[width=\columnwidth,angle=0]{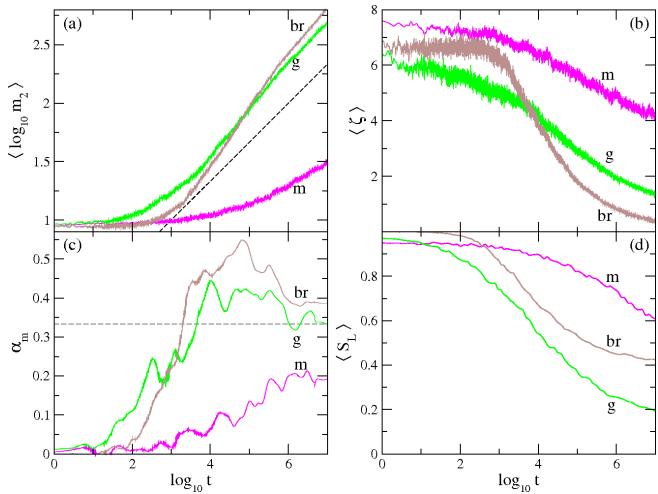} 
\caption{(Color online) DNLS, $W=15$: Evolution of (a) $\langle
  \log_{10}m_2(t) \rangle$, (b) $\langle \zeta(t) \rangle$, (c)
  $\alpha_m (t)$, and (d) $\langle S_L(t) \rangle$ versus $\log_{10}t$
  for the spreading of initially compact wave packets of width $L=10$
  with $\beta=0.5, \, 9 , \, 30$ [(m) magenta; (g) green; (br)
  brown]. Mean values are averaged quantities over 100 disorder
  realizations. In panels (a), and (c) straight lines correspond to
  theoretically predicted weak chaos behavior $m_2 \sim t^{1/3}$. }
\label{fig_DNLS_W_15}
\end{figure}

\begin{figure}
\includegraphics[width=\columnwidth,angle=0]{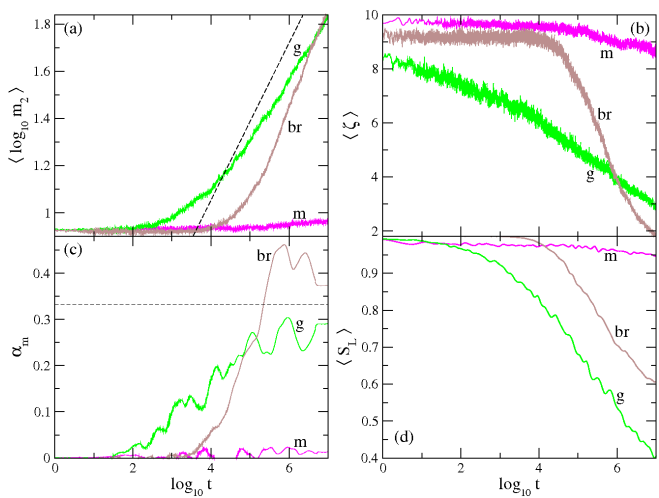} 
\caption{(Color online) DNLS, $W=40$: Evolution of (a) $\langle
  \log_{10}m_2(t) \rangle$, (b) $\langle \zeta(t) \rangle$, (c)
  $\alpha_m (t)$, and (d) $\langle S_L(t) \rangle$ versus $\log_{10}t$
  for the spreading of initially compact wave packets of width $L=10$
  with $\beta=1, \, 25 , \, 100$ [(m) magenta; (g) green; (br)
  brown]. Mean values are averaged quantities over 100 disorder
  realizations. In panels (a), and (c) straight lines correspond to
  theoretically predicted weak chaos behavior $m_2 \sim t^{1/3}$.}
\label{fig_DNLS_W_40}
\end{figure}

To search for potential deviations from the predicted spreading laws,
we turn to large values of $W$ for the DNLS system. In all our
simulations we had $L \ge V$.  In particular, we set $L=10$ and
considered the cases with $W=15$ and $W=40$.  For large values of $W$
we expect only the weak chaos and the selftrapping dynamical regimes
to be observed \cite{LBKSF10}.  For each value of $W$ three different
values of $\beta$ were considered, one being in the weak chaos regime
and the other two in the selftrapping regime. In particular, we
considered $\beta=0.5$, $\beta=9$ and $\beta=30$ for $W=15$ and
$\beta=1$, $\beta=25$ and $\beta=100$ for $W=40$. The obtained results
are shown in Fig.~\ref{fig_DNLS_W_15} for $W=15$ and in
Fig.~\ref{fig_DNLS_W_40} for $W=40$.

For large $W$, the localization volume $V$ decreases drastically, so
that eventually the overlap integrals become small as well.
Therefore, the values of $\beta$ at which spreading becomes visible
will increase.  In the weak chaos spreading regime the detrapping time
$t_d$ becomes large and we start observing spreading after long time
intervals. For $W=15$ and $\beta=0.5$ we find $t_d \approx 10^3$
(Fig.~\ref{fig_DNLS_W_15}(a)).  The local derivative $\alpha_m$
increases from zero showing a tendency to approach the theoretically
predicted value $\alpha=1/3$ (Fig.~\ref{fig_DNLS_W_15}(c)), and both
$\langle \zeta \rangle$ (Fig.~\ref{fig_DNLS_W_15}(b)) and $\langle S_L
\rangle$ (Fig.~\ref{fig_DNLS_W_15}(d)) start to decrease. We note that
since $V\sim 1$ for large $W$, we measure the time evolution of the
fraction $S_L(t)$ of the norm density of the $L=10$ initially excited
sites.

The detrapping time $t_d$ increases as $W$ increases. This is seen
from the results for $W=40$, $\beta=1$ (magenta curves in
Fig.~\ref{fig_DNLS_W_40}). In this case, we have to wait at least up
to $t_d=10^5$ in order to get some evidence that spreading starts,
since after that time $\langle \log_{10} m_2 \rangle$ starts to
slightly grow (Fig.~\ref{fig_DNLS_W_40}(a)), while $\langle \zeta
\rangle$ (Fig.~\ref{fig_DNLS_W_40}(b)) and $\langle S_L \rangle$
(Fig.~\ref{fig_DNLS_W_40}(d)) start to decrease.  This increase of
$t_d$ happens despite the fact that the nonlinearity strength $\beta$
also increased by a factor of two as compared to the $W=15$ case.
Nevertheless even in this case of large $W=40$ we are able to
numerically observe the onset of spreading in the weak chaos
regime. Increasing $W$ to even higher values pushes $t_d$ to values
larger than the final integration time $t=10^7$ used in our
simulations.

With increasing $\beta$ we observe selftrapping, but a part of the
wave packet spreads and $m_2$ increases (green and brown curves in
Figs.~\ref{fig_DNLS_W_15}(a) and \ref{fig_DNLS_W_40}(a)). The average
compactness index $\langle \zeta \rangle$ decreases, which is a clear
indication that a part of the wave packet remains localized, and
reaches smaller final values for larger $\beta$
(Figs.~\ref{fig_DNLS_W_15}(b) and \ref{fig_DNLS_W_40}(b)). The
selftrapping of the wave packets is also clearly seen from the
evolution of $\langle S_L \rangle$. In Fig.~\ref{fig_DNLS_W_15}(d) we
see that for $W=15$, $\beta=9$ (green curve) and $\beta=30$ (brown
curve) $\langle S_L \rangle$ decreases due to the spreading of a part
of wave packets, while, finally it shows a tendency to level off to a
positive value, indicating that part of the wave packets remains
localized. Similar behaviors of $\langle S_L \rangle$ are observed in
Fig.~\ref{fig_DNLS_W_40}(d) for the $W=40$ case with $\beta=25$ (green
curve) and $\beta=100$ (brown curve), although the plateauing of
$\langle S_L \rangle$ is not as clear as in
Fig.~\ref{fig_DNLS_W_15}(d). The numerically computed exponents
$\alpha_m$ exhibit the typical behavior of selftrapping seen in
Fig.~\ref{fig_DNLS_4}(d). For $W=15$ they increase reaching values
larger than $1/3$ and afterwards decrease towards $\alpha_m=1/3$
(Fig.~\ref{fig_DNLS_W_15}(c)). For $W=40$ a similar behavior is
observed for $\beta=100$, while for $\beta=25$ $\alpha_m$ seems to
approach the theoretically predicted value $1/3$ from below.

Therefore, even for strong disorder, the dynamics of wave packets
evolves according to the theoretical predictions. Most importantly, we
do not observe a slowing down of the wave packet below the limits set
by the weak chaos regime.

\subsubsection{FSW model}
\label{sec:FSW}

\begin{figure}
\includegraphics[angle=0,width=0.99\columnwidth]{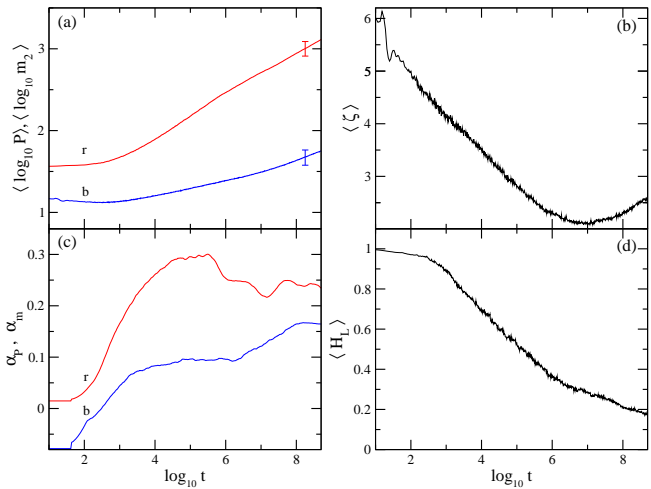}
\caption{(Color online) FSW: Evolution of (a) $\langle \log_{10}m_2(t)
  \rangle$ [(r) red curve] and $\langle \log_{10}P(t) \rangle$ [(b)
  blue curve], (b) $\langle \zeta(t) \rangle$, (c) $\alpha_m (t)$ [(r)
  red curve] and $\alpha_P (t)$ [(b) blue curve], and (f) $\langle
  H_L(t) \rangle$ versus $\log_{10}t$ for initially compact wave
  packets of width $L=21$ with $E=0.05$. }
\label{fig:FSWmodel}
\end{figure}

To further probe a possible slowing down of wave packet spreading
beyond the limits of weak chaos,we turn to the FSW model (\ref{FSW}).
In Fig.~\ref{fig:FSWmodel} we present results with initial compact
wave packets of width $L=21$, with energy density $E=0.05$, similarly
to the KG model (\ref{RQKG}). We observe that also for this model
subdiffusive spreading occurs, because the second moment and the
participation number (red and blue curves respectively in
Fig.~\ref{fig:FSWmodel}(a)) increase continuously, and the
corresponding exponents $\alpha_m$ and $\alpha_P$
(Fig.~\ref{fig:FSWmodel}(c)) tend to eventual constant non-zero
values. The fraction $\langle H_L \rangle$ of energy remaining in the
$L=21$ initially excited sites (Fig.~\ref{fig:FSWmodel}(d)) decreases
as time increases indicating the delocalization of the wave
packets. The compactness index (Fig.~\ref{fig:FSWmodel}(b)) has a
different behavior with respect to what we have seen in the rest of
our simulations, as it decreases slowly up to $t\approx 10^7$, with a
subsequent increase.  Therefore, the wave packet is highly
inhomogeneous for almost all the integration time, violating the
assumptions which are used in the theoretical considerations of weak
chaos \cite{F10}.  The observed subdiffusive spreading may still not
be in its final asymptotic range. Still, we again do not see any
signature of a slowing down of this subdiffusive process, as was
reported in \cite{MAP11} where a similar model was considered. Clearly
the FSW model calls for a thorough and independent study.

\subsubsection{LNL and NLN models}
\label{sec:hybrid}

\begin{figure*}
\includegraphics[scale=0.35,angle=0]{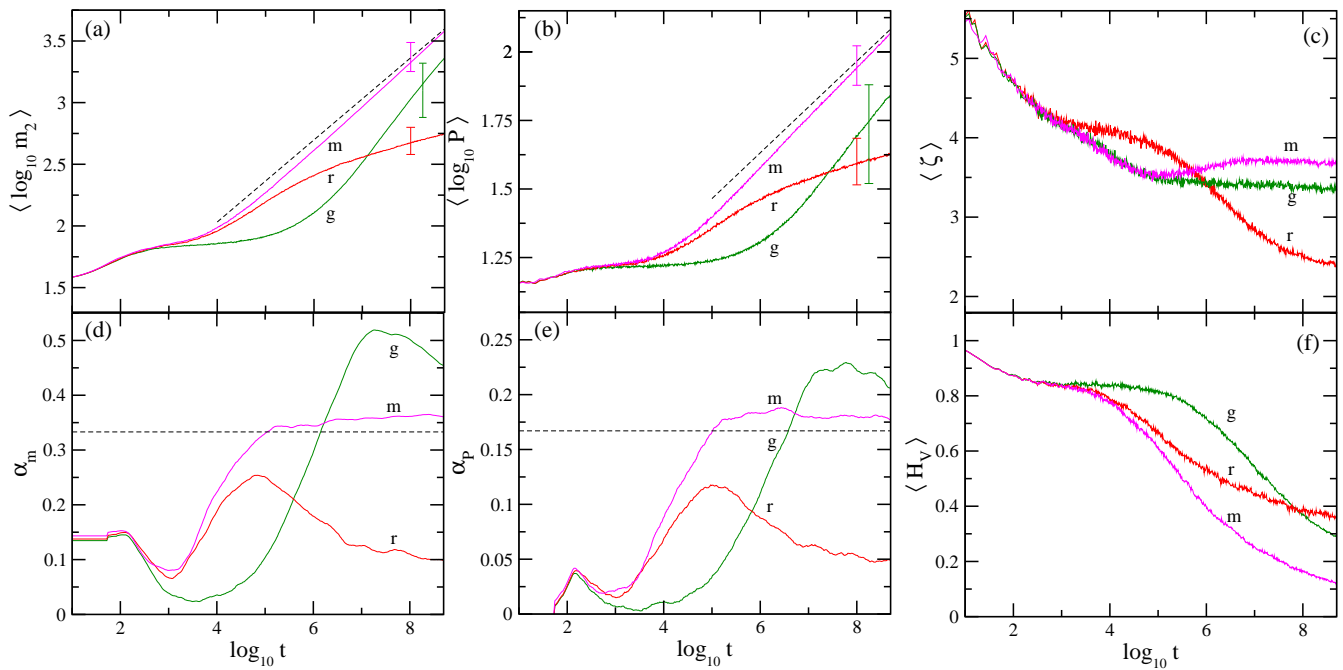} 
\caption{(Color online) Evolution of (a) $\langle \log_{10}m_2(t)
  \rangle$, (b) $\langle \log_{10}P(t) \rangle$, (c) $\langle \zeta(t)
  \rangle$ , (d) $\alpha_m (t)$, (e) $\alpha_P (t)$, and (f) $\langle
  H_V(t) \rangle$ versus $\log_{10}t$ for the spreading of initially
  compact wave packets of width $L=21$ with $W=4$ and $E=0.02$ in the
  KG [(m) magenta curves], the KG-NLN [(g) green curves] and the
  KG-LNL [(r) red curves] models. In panels (a), (b), (d) and (e)
  straight lines correspond to theoretically predicted power laws $m_2
  \sim t^{1/3}$, $P\sim t^{1/6}$ of the weak chaos regime.  Error bars
  in panels (a) and (b) denote representative one standard deviation
  errors.}
\label{fig_KG_hybrid}
\end{figure*}

In Fig.~\ref{fig_KG_hybrid} we present results for the KG, LNL and NLN
models for $W=4$, $L=21$ and $E=0.02$. For comparison we also include
the results for the KG model (\ref{RQKG}) (magenta curves) with
$E=0.02$, for which subdiffusive spreading in the weak chaos regime is
observed. The KG-NLN model (green curves in Fig.~\ref{fig_KG_hybrid})
exhibits a similar behavior, since both the second moment
(Fig.~\ref{fig_KG_hybrid}(a)) and the participation number
(Fig.~\ref{fig_KG_hybrid}(b)) start to grow after some detrapping time
$t_d\approx 10^5$. This time is larger than the detrapping time of the
KG model ($t_d\approx 10^4$), because a wave packet in the KG-NLN
model initially evolves in an almost linear system and only after some
large time, when it has spread significantly to the nonlinear part of
the lattice, spreading takes on characteristics of the purely
nonlinear model.

On the other hand the evolution of all quantities of
Fig.~\ref{fig_KG_hybrid} for the KG-LNL system (red curves) follows
the KG model until $t \approx 10^4$, because initially the wave
packets evolve in the same nonlinear system.  Later on the wave packet
enters the L (linear) parts of the system.  Thus, spreading starts to
retard, and both $\langle \log_{10}m_2(t) \rangle$
(Fig.~\ref{fig_KG_hybrid}(a)) and $\langle \log_{10}P(t) \rangle$
(Fig.~\ref{fig_KG_hybrid}(b)) show a characteristic slowing down in
the exponents $\alpha_m$ (Fig.~\ref{fig_KG_hybrid}(d)) and $\alpha_P$
(Fig.~\ref{fig_KG_hybrid}(e)). In addition, $\langle S_V \rangle$
(Fig.~\ref{fig_KG_hybrid}(f)) saturates at finite non-zero values,
indicating that wave packets tend to localize again.  For all three KG
models, the values of $\langle \zeta \rangle$
(Fig.~\ref{fig_KG_hybrid}(c)) show that wave packets do not become
sparse and inhomogeneous in the course of time.  We obtained similar
results for the LNL and NLN DNLS models for $W=4$, $L=21$ and
$\beta=0.04$.

Thus, also for the LNL and NLN models spreading is observed, and only
in the case of the LNL models we have observed a slowing down of the
spreading, as expected.

\section{Summary and conclusions}
\label{sec:concl}

We considered several models of disordered nonlinear one-dimensional
lattices and performed extensive numerical simulations of norm
(energy) propagations. Since we focused on the dynamical spreading of
fronts, we prepared initial block wave packet profiles, having widths
equal or larger than the average localization volume defined by the
linear problem. While not performed, we expect similar behaviors for
initial Gaussian profiles, again where the width (for Gaussians, say
the standard deviation) is on par with the average localization
volume.

We carefully studied statistical properties of the dynamics, by
varying the values of disorder and nonlinearity strengths over a wide
interval, and by averaging results over many disorder
realizations. Our results agree quite well with our theoretical
expectations for the existence of the weak and strong chaos regimes.

The main outcome of our study is that in the presence of
nonlinearities we always observe subdiffusive spreading, so that the
second moment grows initially as $m_2 \sim t^\alpha$ with $\alpha<1$,
showing signs of a crossover to the asymptotic $m_2 \sim t^{1/3}$ law
at larger times. Remarkably, subdiffusive spreading is also observed
for large disorder strengths, when the localization volume (which
defines the number of interacting partner modes) tends to one.
Fr\"ohlich-Spencer-Wayne models which take the disorder strength to
its infinite limits, are also showing subdiffusive growth.  Most
remarkably, in none of our studies (except the artificial LNL case)
did we encounter a slowing down of spreading beyond the limits set by
the weak chaos predictions. Therefore, our numerical data support the
conjecture, that the wave packets, once they spread, will do so up to
infinite times in a subdiffusive way, bypassing Anderson localization
of the linear wave equations.

The only cases where spreading shows a tendency to stop are the LNL
models, for which nonlinearities are absent everywhere except inside a
finite-size central region, where the initial wave packet is
launched. In these models, when wave packets have spread
substantially, their chaotic component in the central region of the
lattice becomes weak, and distant normal modes in the linear parts of
the system are exponentially weakly coupled to the central nonlinear
region.

When the nonlinearity strength tends to smaller values, waiting
(detrapping) times for wave packet spreading of compact initial
excitations increase beyond the detection capabilities of our
computational tools. The corresponding question of whether a KAM
regime can be entered at finite nonlinearity strength was addressed in
\cite{JKA10} and is analyzed in detail in a forthcoming work
\cite{ILF11}.

\begin{acknowledgments}
  We thank S.~Aubry, M.~Mulansky, A.~Pikovsky, R. Schilling and
  D.~Shepelyansky for useful discussions. Ch.~S.~was partly supported
  by the European research project ``Complex Matter'', funded by the
  GSRT of the Ministry Education of Greece under the ERA-Network
  Complexity Program.
\end{acknowledgments}

\appendix

\section{Symplectic integration of the DNLS equations}
\label{sec:DNL_integration}

We first discuss a novel PQ method which we designed to integrate the
DNLS equations locally. Previously used methods employ a
transformation of the wave function from real into Fourier space and
back, at each integration step. These transformations induce small but
observable corrections in the tails of the wave packet, which slowly
but steadily grow in time.  In such a case we will have to stop the
integration once this noisy background reaches a substantial
level. The PQ method avoids the generation of this background by
simply not performing the Fourier transformation. Instead the PQ
method integrates the DNLS equations in real space.

The canonical transformation 
\begin{eqnarray}
\psi_l=\dfrac{1}{\sqrt{2}}(q_l+ip_l)
\label{eq:trans}
\end{eqnarray}
of the complex variable $\psi_l$ in Eq.~(\ref{RDNLS}) transforms (\ref{RDNLS})
into
\begin{equation}
{\cal
H}_D=\sum_l\dfrac{\epsilon_l}{2}(q_l^2+p_l^2)+\dfrac{\beta}{8}(q_l^2+p_l^2)^2-
(q_{l+1}q_l+ p_{l+1} p_l).
\label{RDNLS_PQ}
\end{equation}
where $q_l$ and $p_l$ are generalized coordinates and momenta,
respectively.

If a Hamiltonian function can be split into two integrable parts, then
a symplectic integration scheme can be used for the integration of its
equations of motion. One possible splitting of the DNLS Hamiltonian
(\ref{RDNLS_PQ}) into two separate Hamiltonian functions $A$ and $B$
is
\begin{equation}
\begin{array}{lll}
  A & = & \displaystyle
  \sum_l\dfrac{\epsilon_l}{2}(q_l^2+p_l^2)+\dfrac{\beta}{8}(q_l^2+p_l^2)^2, \\
  B & = & \displaystyle -\sum_l(q_{l+1}q_l+ p_{l+1} p_l).
\end{array}
\label{eq:ham_equiv_DNLS}
\end{equation}
Hamiltonian $A$ is integrable and the operator $e^{\tau L_A}$ which
propagates the set of initial conditions $(q_l, p_l)$ at time $t$, to
their final values $(q'_l, p'_l)$ at time $t+\tau$ is
\begin{equation}
e^{\tau L_A}: \left\{ \begin{array}{lll} q'_l & = & q_l \cos(\alpha_l \tau)+ p_l
\sin(\alpha_l \tau)\\ p'_l & =& p_l \cos(\alpha_l \tau)- q_l \sin(\alpha_l
\tau) \\
\end{array}\right. ,
\label{eq:LA}
\end{equation}
with $\alpha_l=\epsilon_l+\beta(q_l^2+p_l^2)/2$.  Hamiltonian $B$ of
Eq.~(\ref{eq:ham_equiv_DNLS}) is not integrable, thus the operator
$e^{\tau L_B}$ cannot be written explicitly. If we consider $B$ as a
separate Hamiltonian function and again split it as $B=P+Q$, the
component parts
\begin{equation}
P=-\sum_lp_{l+1} p_l, \,\, Q= -\sum_lq_{l+1}q_l.
\label{eq:PQ}
\end{equation}
 are integrable, under the corresponding operators
\begin{equation}
e^{\tau L_P}: \left\{ \begin{array}{lll} p'_l & =& p_l \\ q'_l & = &
q_l-(p_{l-1}+p_{l+1}) \tau \\
\end{array}\right. ,
\label{eq:LP}
\end{equation}
and
\begin{equation}
e^{\tau L_Q}: \left\{ \begin{array}{lll} q'_l & =& q_l \\ p'_l & = &
p_l+(q_{l-1}+q_{l+1}) \tau \\
\end{array}\right. 
\label{eq:LQ}
\end{equation}
This technique of splitting the Hamiltonian into multiple parts has
been used in different applications of symplectic integrators (see for
example \cite{GBB08}).

In our simulations we successively apply the SBAB$_2$ symplectic
integrator \cite{LR01,SKKF09,SG10} twice: first for the split (${\cal
  H}_D=A+B$) of the DNLS Hamiltonian, Eq.~(\ref{RDNLS_PQ}), and second
for the split $B=P+Q$ in Eq.~(\ref{eq:ham_equiv_DNLS}).  The solution
for the equations of motion from the Hamiltonian Eq.~(\ref{RDNLS_PQ})
is thus approximated by the application of $13$ simple operators on an
initial condition $(q_l, p_l)$, since
\begin{widetext}
\begin{equation}
\begin{array}{lll}
\displaystyle e^{\tau {\cal H}_D} &\displaystyle = & \displaystyle
e^{\tau(A+B)} \approx e^{d_1\tau L_A}e^{c_2\tau L_B}e^{d_2\tau L_A}e^{c_2\tau
L_B}e^{d_1\tau L_A} \approx \\ &\displaystyle \approx & \displaystyle e^{d_1\tau
L_A}e^{d_1c_2\tau L_P}e^{c_2^2\tau L_Q}e^{c_2 d_2\tau L_P}e^{c_2^2\tau
L_Q}e^{d_1 c_2\tau L_P}e^{d_2\tau L_A} e^{d_1c_2\tau L_P}e^{c_2^2\tau L_Q}e^{c_2
d_2\tau L_P}e^{c_2^2\tau L_Q}e^{d_1 c_2\tau L_P}e^{d_1\tau L_A},
\end{array}
\label{eq:DNLSPQ}
\end{equation}
\end{widetext}
with the SBAB$_2$ coefficients~\cite{LR01} of $d_1=1/6$, $d_2=2/3$ and $c_2=1/2$.


\end{document}